\documentclass[prd, aps, superscriptaddress, preprintnumbers, twocolumn, floatfix, nofootinbib]{revtex4}
\pdfoutput=1

\usepackage{amsfonts}
\usepackage{amsmath}
\usepackage{amssymb}
\usepackage{bm}
\usepackage{dcolumn}
\usepackage{graphicx}   
\usepackage[latin1]{inputenc}
\usepackage{latexsym}
\usepackage{rotating}
\usepackage{hyperref}
\usepackage{graphicx}
\usepackage{color}
\usepackage{comment}
\usepackage{ulem}
\usepackage{url}

\usepackage{natbib}
\usepackage{comment}
\usepackage{xcolor}

\begin{document}

\title{SKYFAST: Optimizing gravitational wave host localization with rapid volume reconstruction and inclination angles}

\title{SKYFAST: Rapid volume reconstruction and inclination angle for gravitational wave host localization}

\title{Enhancing gravitational wave host localization with SKYFAST: rapid volume reconstruction and inclination angle}

\title{Enhancing gravitational-wave host localization with \texttt{SKYFAST}:\\ rapid volume and inclination angle reconstruction}

\author{Gabriele Demasi}
\email{gabriele.demasi@unifi.it}
\affiliation{Dipartimento di Fisica e Astronomia, Universit\`a degli Studi di Firenze, \\ Via Sansone 1, Sesto Fiorentino (Firenze) I-50019, Italy}
\affiliation{INFN, Sezione di Firenze, Sesto Fiorentino (Firenze) I-50019, Italy}

\author{Giulia Capurri}
\email{giulia.capurri@df.unipi.it}
\affiliation{Dipartimento di Fisica ``Enrico Fermi'', Universit\`a di Pisa, \\ Largo Bruno Pontecorvo 3, Pisa I-56127, Italy}
\affiliation{INFN, Sezione di Pisa, Largo Bruno Pontecorvo 3, Pisa I-56127, Italy}

\author{Angelo Ricciardone}
\affiliation{Dipartimento di Fisica ``Enrico Fermi'', Universit\`a di Pisa, \\ Largo Bruno Pontecorvo 3, Pisa I-56127, Italy}
\affiliation{INFN, Sezione di Pisa, Largo Bruno Pontecorvo 3, Pisa I-56127, Italy}
\affiliation{Dipartimento di Fisica e Astronomia ``Galileo Galilei'',\\ Universit\`a di Padova, I-35131 Padova, Italy}

\author{Barbara Patricelli}
\affiliation{Dipartimento di Fisica ``Enrico Fermi'', Universit\`a di Pisa, \\ Largo Bruno Pontecorvo 3, Pisa I-56127, Italy}
\affiliation{INFN, Sezione di Pisa, Largo Bruno Pontecorvo 3, Pisa I-56127, Italy}
\affiliation{INAF - Osservatorio Astronomico di Roma, \\ Via Frascati 33, Monte Porzio Catone (Rome) I-00078, Italy}

\author{Massimo Lenti}
\affiliation{Dipartimento di Fisica e Astronomia, Universit\`a degli Studi di Firenze, \\ Via Sansone 1, Sesto Fiorentino (Firenze) I-50019, Italy}
\affiliation{INFN, Sezione di Firenze, Sesto Fiorentino (Firenze) I-50019, Italy}

\author{Walter Del Pozzo}
\affiliation{Dipartimento di Fisica ``Enrico Fermi'', Universit\`a di Pisa, \\ Largo Bruno Pontecorvo 3, Pisa I-56127, Italy}
\affiliation{INFN, Sezione di Pisa, Largo Bruno Pontecorvo 3, Pisa I-56127, Italy}

\date{\today}

\begin{abstract}
The scientific impact of GW170817 strongly supports the need for an efficient electromagnetic follow-up campaign to gravitational-wave event candidates. The success of such campaigns depends critically on a fast and accurate localization of the source. 
In this paper, we present \texttt{SKYFAST}, a new pipeline for rapid localization of gravitational-wave event hosts. \texttt{SKYFAST} runs alongside a full parameter estimation (PE) algorithm, from which posterior samples are taken. It uses these samples to reconstruct an analytical posterior for the sky position, luminosity distance, and inclination angle using a Dirichlet Process Gaussian Mixture Model, a Bayesian non-parametric method. This approach allows us to provide an accurate localization of the event using only a fraction of the total samples produced by the full PE analysis. Depending on the PE algorithm employed, this can lead to significant time savings, which is crucial for identifying the electromagnetic counterpart.
Additionally, in a few minutes, \texttt{SKYFAST} generates a ranked list of the most probable galaxy hosts from a galaxy catalog of choice. This list includes information on the inclination angle posterior conditioned to the position of each candidate host, which is useful for assessing the detectability of gamma-ray burst structured jet emissions.

\end{abstract}

\maketitle

\section{Introduction}

The era of multi-messenger astronomy with gravitational waves (GWs) began on August 17, 2017, with the detection of the binary neutron star (BNS) merger GW170817 \cite{LIGOScientific:2017vwq} by the Advanced LIGO \cite{LIGOScientific:2014pky} and Advanced Virgo \cite{VIRGO:2014yos} interferometers . 
This was immediately followed by the independent observation of the short gamma-ray burst (GRB) GRB 170817A by the Gamma-ray Burst Monitor onboard \textit{Fermi} \cite{Goldstein:2017mmi} and the Spectrometer Anti-Coincidence Shield onboard the International Gamma-Ray Astrophysics Laboratory (INTEGRAL) \cite{Savchenko:2017ffs}. 
An intense electromagnetic (EM) follow-up campaign, was launched across the entire EM spectrum after 
this joint detection, and especially after 
the subsequent well-constrained, three-dimensional LIGO-Virgo source localization
(see \cite{LIGOScientific:2017ync} and references therein). 
The campaign led to the discovery of an optical transient source, SSS17a/AT 2017gfo, located in the NGC 4993 galaxy \cite{Coulter:2017wya}. 
This source, observed also in ultraviolet and infrared wavelengths, was then identified as a kilonova \cite{Cowperthwaite:2017dyu,Evans:2017mmy,Arcavi:2017xiz}. The ensemble of all the GW and EM measurements associated with this event constitutes the first multi-messenger observation with GWs. Such a discovery led to a wide range of extraordinary results, including the identification of BNS mergers as progenitors of short GRBs, the evidence of a structure in GRB jets, the precise measurement of the GW propagation speed, and a new independent measurement of the Hubble constant \cite{Mooley:2018clx,Ghirlanda:2018uyx, LIGOScientific:2017zic,LIGOScientific:2017adf}.

GW170817 remains the only confident multi-messenger GW event to date (see, however, \cite{Graham:2020gwr, Connaughton:2016umz, Ashton:2020kyr, 
 Morton:2023wxg} and references therein), but we expect more multi-messenger events in the future as the sensitivity of GW facilities improves, see, e.g., \cite{Patricelli:2022hhr,Stratta:2022jaj,Corsi:2024vvr,Colombo:2022zzp,Colombo:2023une,Bhattacharjee:2024wyz}. An EM counterpart is expected from the merger of binary systems containing at least one neutron star, i.e., BNSs and neutron star-black hole (NS-BH) binaries \cite{Lattimer:1974slx,Eichler:1989ve}. After the merger and subsequent GW generation, the GRB occurs, consisting of a prompt $\gamma$-ray emission the duration of which is $\lesssim 2$ s, and a multi-wavelength afterglow typically observable in the optical, X-ray, and radio bands for minutes, hours, days, and even months after the prompt emission. The optical/near-infrared transient associated with the kilonova is expected to appear hours/days after the prompt emission and progressively fade away.
In this context, it is clear that a rapid and precise localization of the GW event is crucial to launch an effective EM follow-up campaign. Currently, when a GW event is detected, it takes from seconds to minutes to have an initial estimate of the event localization thanks to \texttt{BAYESTAR} \cite{Singer:2015ema}, a Bayesian, non-Markov Chain Monte Carlo (MCMC) algorithm for the rapid localization of GW events. The initial \texttt{BAYESTAR} volume maps are released to the astronomical community so that the hunt for the EM counterpart can promptly start \cite{Singer:2016eax,Singer:2016erz}. However, to achieve a definitive, more precise localization of the GW event, one must wait for the results of a full parameter estimation (PE) pipeline, which generally takes longer due to the large number of parameters whose posterior must be reconstructed as well as to the complexity of the sampling algorithms used to explore the likelihood.

In this paper, we present \texttt{SKYFAST}, a new localization algorithm based on Bayesian non-parametrics, extending the idea introduced in \cite{Rinaldi:2022kyg}.
\texttt{SKYFAST} takes as input the samples produced by a PE pipeline (e.g., \texttt{LALInference} \cite{Veitch:2014wba}, \texttt{Bilby} \cite{Ashton:2018jfp,Romero-Shaw:2020owr, morisaki:2023}, \texttt{JIM} \cite{Wong:2023lgb,Wouters:2024oxj}) and uses them to reconstruct an analytic posterior distribution for the sky position, luminosity distance, and inclination angle of the GW event. The posterior is modelled as a weighted sum of multivariate Gaussians, whose weights, means, and covariance matrices are outcomes of a Dirichlet Process Gaussian Mixture Model (DPGMM) \cite{DelPozzo:2018dpu,Rinaldi:2021bhm}. Specifically, we use the DPGMM implementation \texttt{FIGARO} \cite{Rinaldi:2022kyg, Rinaldi:2024eep}. In \texttt{SKYFAST}, the Gaussian mixture is progressively updated as new samples are produced by the PE algorithm until the information entropy~\cite{Shannon:1948nms} reaches a plateau, indicating that the addition of more samples would not improve the quality of the reconstructed posterior. At this point, the accuracy of the posterior is comparable to the final PE results. This convergence criterion is typically reached in a fraction of the PE runtime, allowing \texttt{SKYFAST} to release an intermediate, yet accurate and fully analytical, joint posterior distribution of the sky localization, luminosity distance, and inclination angle ahead of time.

Together with the reconstructed posterior, \texttt{SKYFAST} releases a skymap of the GW event and a ranked list of the most probable hosts from a galaxy catalog of choice. In this paper, we use the GLADE+ catalog \cite{Dalya:2021ewn} as an example to display the great potential of \texttt{SKYFAST}, but the algorithm can be easily adapted to work with any galaxy catalog. The ranked list includes median and 90\% credible intervals for the inclination angle posterior conditioned on the position of each galaxy. This information allows for specific modeling of the GRB structured jet emission based on the location of the event within the identified potential host galaxies. 
The inclusion of the inclination angle is a novel feature of \texttt{SKYFAST}. Currently, the LIGO-Virgo-KAGRA (LVK) Collaboration does not release the inclination angle posterior in the alerts for EM follow-up campaigns. We show that the inclination angle information provided by \texttt{SKYFAST} can be relevant for optimizing the follow-up tiling strategies and tailoring the efforts to each candidate host, under the hypothesis that the GW event originated in that particular galaxy. 
  
The features of the different PE algorithms determine how \texttt{SKYFAST} can be integrated into the pipeline, its speed performance, and which of its outputs can be used. Most importantly, using \texttt{SKYFAST} to produce an intermediate analytical posterior reconstruction is only possible if the PE algorithm releases the samples during the analysis run, as is the case for MCMC samplers. This is not possible with nested sampling algorithms, where posterior samples are only available at the end of the run. Nonetheless, \texttt{SKYFAST} can always be used on the final set of samples to produce a list of the most probable galaxy hosts. 
\texttt{SKYFAST} is publicly available at \url{https://github.com/gabrieledemasi/skyfast}.\\
The paper is structured as follows: in Sec. \ref{sec: ex_summary}, we outline the main features and outputs of \texttt{SKYFAST}; in Sec. \ref{sec:validation}, we validate its statistical robustness on a mock dataset of GW signals; in Sec. \ref{sec:discussion}, we discuss the combination between \texttt{SKYFAST} and different PE algorithms, as well as the role \texttt{SKYFAST} can play in optimizing EM follow-up searches;
finally, we draw our conclusions in Sec.\ref{sec:conclusion}.

\section{Executive summary}\label{sec: ex_summary}
\texttt{SKYFAST} is a tool designed for the rapid reconstruction of the joint posterior of the localization parameters, namely, right ascension ($\alpha$), declination ($\delta$), luminosity distance ($d_\text{L}$), and the inclination angle ($\theta_{jn}$), which is defined as the angle between the direction of observation and the total angular momentum of the binary.
For the reconstruction of the localization parameters, we follow the procedure discussed in \cite{Rinaldi:2022kyg}. A key difference, however, is that \texttt{SKYFAST} processes the posterior samples one by one, so that it can effectively work in parallel with any PE pipeline that releases the samples during the run. Moreover, differently from \cite{Rinaldi:2022kyg} and from existing galaxy ranking tools (e.g., \cite{Salmon:2019bqy, ned_gwf,Arcavi:2017vbi, Ducoin:2019rdv}) we also include the reconstruction of the inclination angle. Indeed, the inclination angle is a key ingredient to estimate the expected temporal evolution of the multi-wavelength flux associated with possible EM counterparts, such as GRB structured jet emission \cite{Ryan:2019fhz}, and therefore to evaluate their detectability.

\texttt{SKYFAST} takes the posterior samples produced by a PE pipeline as input and reconstructs an analytical posterior for the parameters $\{ \alpha, \delta, d_\text{L}, \theta_{jn} \}$ using \texttt{FIGARO} \cite{Rinaldi:2024eep}, a publicly available\footnote{\url{https://github.com/sterinaldi/FIGARO.git}} inference code designed to estimate multivariate probability distributions using a DPGMM. A DPGMM is a non-parametric Bayesian method that, given a set of samples, reconstructs the distribution from which they have been drawn as a mixture of multivariate Gaussians with means $\bm{\mu}_k$ and covariances $\bm{\sigma}_k$:

\begin{equation}
    p(\textbf{x}) \simeq \sum_{k=1}^{N_{G}} w_k \, \mathcal{N}(\textbf{x}|\bm{\mu}_k,\bm{\sigma}_k)\,,
\end{equation}
where the number of components of the mixture $N_{G}$ and the mixing proportions $w_k$
 are inferred from a Dirichlet process \cite{doi:10.1080/01621459.1995.10476550,NIPS1999_97d98119}. For further details about the DPGMM please refer to \cite{Rinaldi:2021bhm, DelPozzo:2018dpu} 
 
\begin{figure}[t!]
    \centering
    \includegraphics[width =.49 \textwidth]{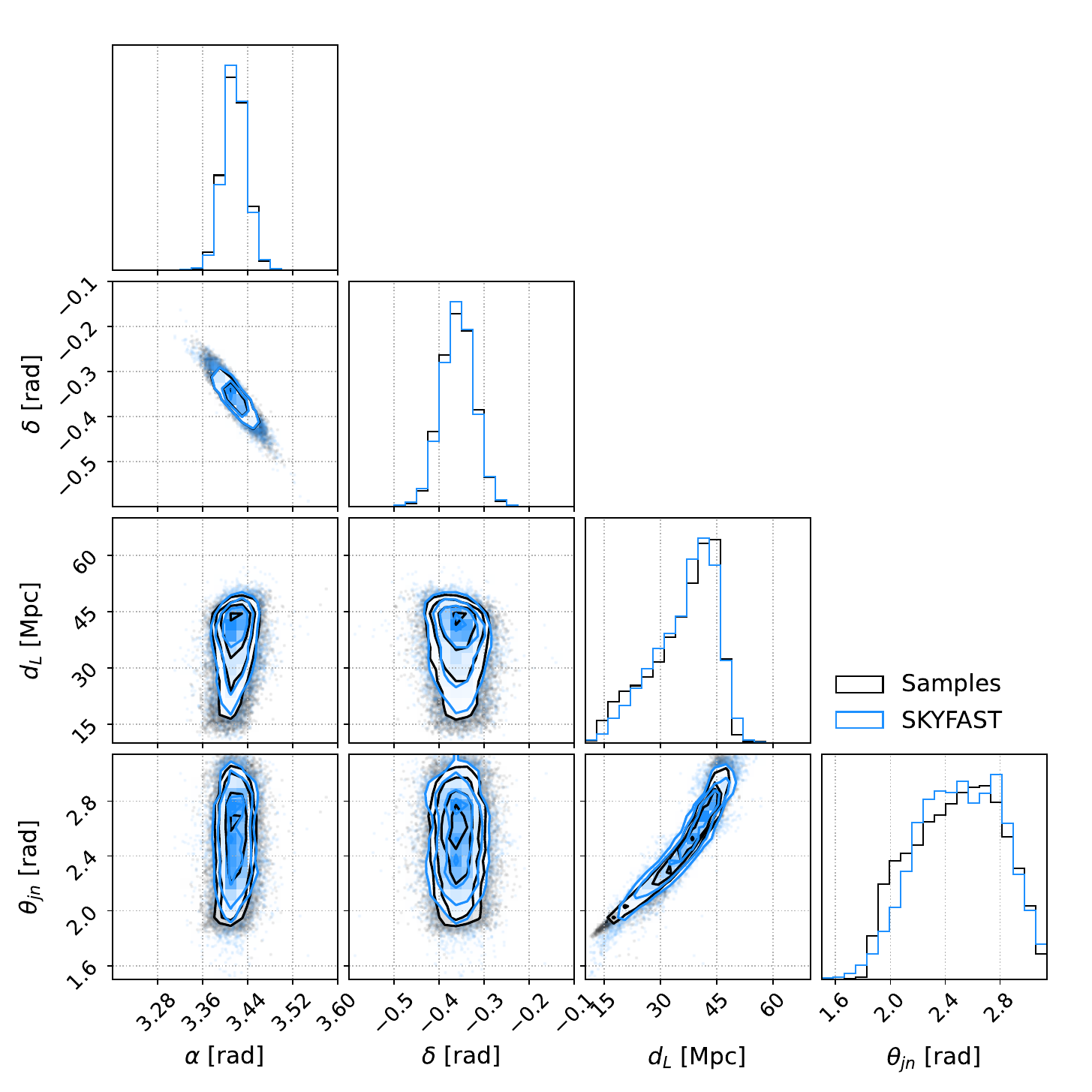}
    \caption{Corner plot of the joint posterior distribution of $\{\alpha, \delta, d_L, \theta_{jn} \}$ for GW170817 . We compare the full PE posterior samples from \cite{Finstad:2018wid} (black) with those obtained from the \texttt{SKYFAST} reconstruction after reaching convergence (blue).}
    \label{fig:corner_plot_gw170817}
\end{figure}

 The mixture is updated continuously as new samples from the PE run are added, until adding further samples no longer provides additional information. We quantitatively estimate the information encoded in the posterior using the information entropy, defined as \cite{Shannon:1948nms}

\begin{equation}
    S(N) = -\sum_k  p_k^{N} \log p_k^{N}\,,
    \label{eq:info_entropy}
\end{equation}
where $p_k^{N}$ is the posterior obtained with $N$ samples, evaluated at the $k$th element of a discrete grid.
Convergence is achieved when $S(N)$ reaches a plateau, and consequently, its derivative begins oscillating around zero. We assume, as in \cite{Rinaldi:2022kyg}, that the reconstruction has converged to its target distribution when the derivative  $dS(N)/dN$  undergoes a given number of zero crossings. Empirical analysis shows that three zero crossings are an effective indicator of convergence. At this point, \texttt{SKYFAST} is ready to release an intermediate reconstructed posterior distribution that already contains the same amount of information as the posterior obtained using all samples.
The posterior obtained with \texttt{SKYFAST} is analytical and, being a mixture of Gaussians, allows for straightforward marginalization and conditioning on any subset of its parameters.\\
In the remainder of this section, we discuss the main outcomes of \texttt{SKYFAST}, using GW170817 as a reference example. Specifically, we use the posterior samples from \cite{Finstad:2018wid}~\footnote{The posterior samples are publicly available at \url{https://github.com/sugwg/gw170817-inclination-angle}}.
The primary output of \texttt{SKYFAST} is the analytical joint posterior for $\{\alpha, \delta, d_L, \theta_{jn}\}$. In Figure \ref{fig:corner_plot_gw170817}, we compare the full PE posterior samples of GW170817 with the samples from posterior distributions reconstructed with \texttt{SKYFAST}. In particular, we use the posterior of the intermediate reconstruction, which is released immediately after reaching convergence.

With an analytical posterior, it is straightforward to evaluate the probability of each galaxy in a given catalog being the host of the GW event. For this work, we use the GLADE+ catalog \cite{Dalya:2021ewn}, which is currently the most complete catalog available for EM follow-up of GW events. However, \texttt{SKYFAST} can be easily adapted to work with any catalog, which could be very useful for the specific requirements of different EM observatories. In its current form, \texttt{SKYFAST} produces a list of all galaxies from the GLADE+ catalog that fall within the 90\% credible volume, ranked by their probability of being the host of the GW event according to their 3D position. Explicitly, the probability of being the host is obtained by evaluating the posterior distribution marginalized over the inclination angle at the position of each galaxy.\footnote{We convert the galaxy's measured redshift to luminosity distance using the Planck18 cosmology \cite{Planck:2018vyg}. } 

\begin{figure}[t]
    \centering
    \includegraphics[width=0.49\textwidth]{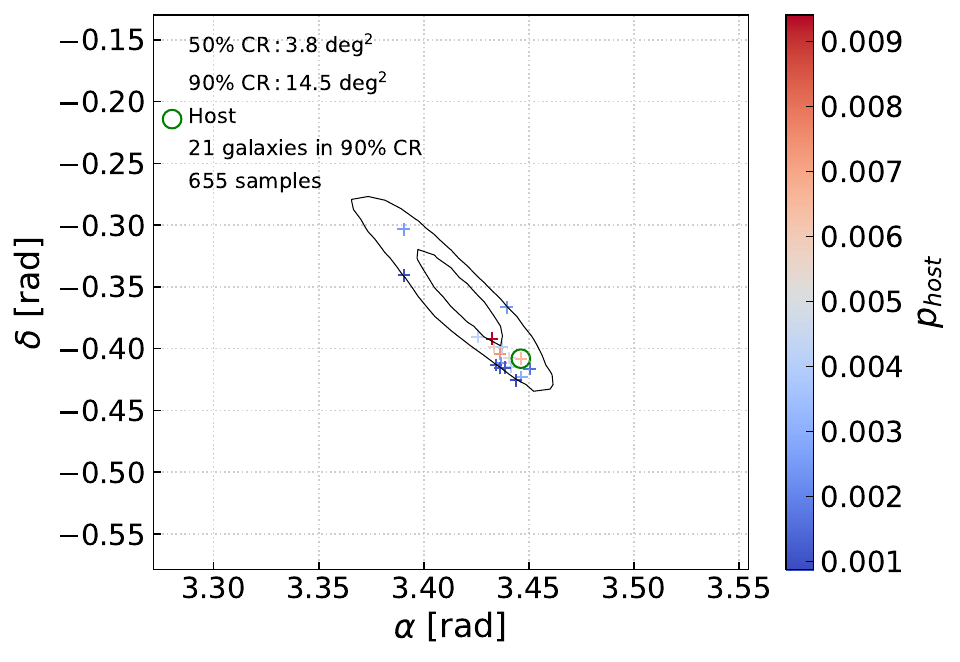}
    \label{fig:image1}
    \caption{Sky localization of GW170817 obtained from the intermediate posterior reconstruction. The solid black lines represent the 50\% and 90\% credible regions. Colored crosses denote the positions of galaxies from the GLADE+ catalog \cite{Dalya:2021ewn} found within the 90\% credible region, ranked by their probability of being the host. The true host, NGC 4993, is marked with a green circle. }
    \label{fig:intermediate_skymap}
\end{figure}

In Figure \ref{fig:intermediate_skymap}, we show the sky localization region of GW170817 obtained using the intermediate posterior distribution, along with all the galaxies from the GLADE+ catalog contained within the 90\% credible volume.
In Table \ref{tab:gw170817_list},  we list the five most probable host galaxies for GW170817, with the true host, NGC4993, in the third position. 

\begin{figure}[t!]
    \centering
    \includegraphics[width = 0.46\textwidth]{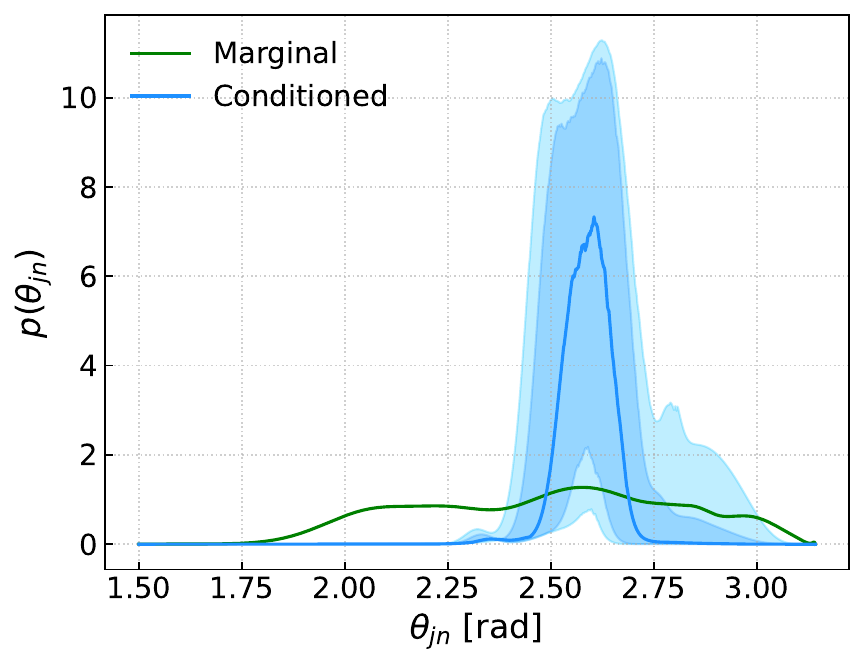}
    \caption{Comparison between the inclination angle posterior distribution marginalized over the position parameters (green) and conditioned to the position of NGC 4993 (blue). The blue line is the median of the distributions obtained by conditioning to 500 different positions extracted from a Gaussian distribution with mean and standard deviation given by the photometric redshift measurement for NGC 4993 reported in the GLADE+ catalog. The shaded areas denote the 68\% and 90\% credible regions.}
    \label{fig:marginal_vs_conditioned}
\end{figure}

We further exploit the analytical reconstruction of the posterior to condition the inclination angle distribution on the position of each galaxy found within the localization region, breaking the degeneracy between $d_L$ and $\theta_{jn}$ \cite{Usman:2018imj}.
Since the redshift measurements for most galaxies in the GLADE+ catalog are photometric, the uncertainty in the luminosity distance is typically of the order of a few percents and, hence, non-negligible. To incorporate this uncertainty in our analysis, we assume a Gaussian distribution for the measured galaxy distances and marginalize over it.
In the last column of Table \ref{tab:gw170817_list}, we report the median value of $\theta_{jn}$ along with 90\% confidence intervals obtained from the posterior distribution conditioned to the position of each galaxy. These intervals encompass the uncertainty in the luminosity distance propagated from the error in the photometric redshift measurements.
In Figure \ref{fig:marginal_vs_conditioned}, we show a comparison between the inclination angle posterior distribution marginalized over the position parameters and conditioned to the position of NGC 4993. The conditioned posterior is computed by averaging over the distributions obtained by conditioning the posterior to 500 different positions drawn from a Gaussian distribution centered on the true NGC 4993 luminosity distance, with standard deviation given by the photometric redshift measurement as reported in the GLADE+ catalog. The shaded areas denote the 68\% and 90\% credible regions. As expected, conditioning to a specific position instead of marginalizing over all position parameters results in a narrower inclination angle posterior. This occurs because fixing the position breaks the degeneracy between $d_{L}$ and $\theta_{jn}$. Remarkably, the median value of $\theta_{jn}$ and its associated errors, obtained from the conditioned posterior for NGC 4993, are consistent with values reported in \cite{Finstad:2018wid}.

\begin{table*}[ht!]
    \centering
    \setlength{\extrarowheight}{5pt}
    \begin{tabular}{cccccccc}
    \hline 
    \textbf{Galaxy} & $\boldsymbol{\alpha}$ [rad] & $\boldsymbol{\delta}$ [rad] & $\boldsymbol{d_{L}}$ [Mpc] & \textbf{logP} & 
    $\boldsymbol{\theta_{jn}}$ [rad] \\[2pt]
    \hline
    ESO 575-055 & 3.43248 & -0.39193 & 45.30372 & -4.18092 & $2.82^{+0.21}_{-0.18}$  \\ 
    ESO 508-014 & 3.44065 & -0.40749 & 44.53725 & -4.51669 & $2.78^{+0.18}_{-0.18}$ \\ 
    NGC 4993 & 3.44613 & -0.40812 & 41.16691 & -4.75818 & $2.63^{+0.20}_{-0.15}$  \\ 
    J131045.95-235156.6 & 3.45037 & -0.41654 & 40.30587 & -5.29308 & $2.59^{+0.19}_{-0.13}$   \\
    PGC 803966 & 3.43619 & -0.40440 & 39.46450 & -5.34645 & $2.56^{+0.16}_{-0.13}$  \\ 
    \hline
    \end{tabular}
    \caption{The five most probable host galaxies for GW170817 from the GLADE+ catalog. The probability of each galaxy being the host is computed according to the intermediate posterior reconstructed with \texttt{SKYFAST}. The last column shows the median value of the inclination angle posterior distribution conditioned to the position of each galaxy. The 90\% credible intervals include the uncertainty on $d_{L}$, propagated from the photometric estimate of the redshift. The true host NGC 4993 is ranked as third.}
    \label{tab:gw170817_list}
\end{table*}

\section{Validation of the algorithm on simulated datasets}\label{sec:validation}

In the following, we test the statistical robustness and performance of \texttt{SKYFAST}. In section \ref{sec:intermediate}, we consider the intermediate reconstruction of the posterior, while in section \ref{sec:galaxyrank}, we show how \texttt{SKYFAST} performs in galaxy ranking.

\subsection{Intermediate posterior reconstruction}\label{sec:intermediate}

We test the robustness of the intermediate reconstruction of the posterior distribution by running \texttt{SKYFAST} on a set of simulated binary black-hole (BBH) events.
We inject 100 BBH signals into Gaussian noise based on the publicly available LVK sensitivity curves\footnote{LVK sensitivity curves publicly available at \url{https://dcc.ligo.org/LIGO-T2000012/public}}. Specifically, we consider a network composed by the two Advanced LIGO detectors and the Advanced Virgo detector. For LIGO, we use the high-noise estimate for the fourth observing run (O4), while for Virgo, we use the power spectral density representative of the third observing run (O3) \cite{KAGRA:2013rdx}.
We use the \texttt{IMRPhenomD} waveform 
\cite{Khan:2015jqa,Husa:2015iqa} both for signal injection and PE. We use the \texttt{Bilby-MCMC} sampler \cite{Ashton:2021anp} implemented in \texttt{Bilby} \cite{Ashton:2018jfp,Romero-Shaw:2020owr}, set to produce 5000 independent samples using 10 parallel runs. Since, with this settings, the median PE runtime is of about one hour, we decide to save the posterior Posterior samples are then extracted from checkpoint files created every five minutes, given the median sampling time of about 1 hour.

\begin{figure}[ht]
    \centering
    \includegraphics[width = 0.45\textwidth]{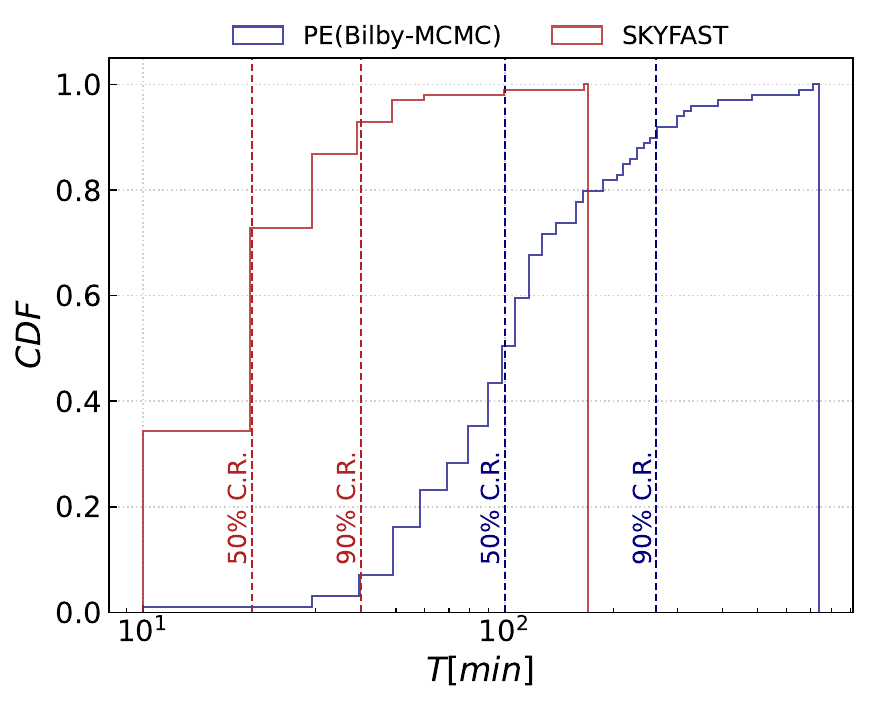}
    \caption{Cumulative distribution of the total PE runtime and the time required by \texttt{SKYFAST} to produce the intermediate posterior reconstruction, the associated skymap and the host galaxy ranking. We obtain the results shown in this plot using the \texttt{Bilby-MCMC} sampler to analyse 100 mock BBH events with the \texttt{IMRPhenomD} waveform. \texttt{SKYFAST} is run on Intel\textsuperscript{\textregistered} Xeon\textsuperscript{\textregistered} Gold 6140M processors with a 2.30 GHz clock rate.}  
    \label{fig:PE/skyfast}
\end{figure}

\begin{figure}[ht]
    \centering
    \includegraphics[width = 0.45\textwidth]{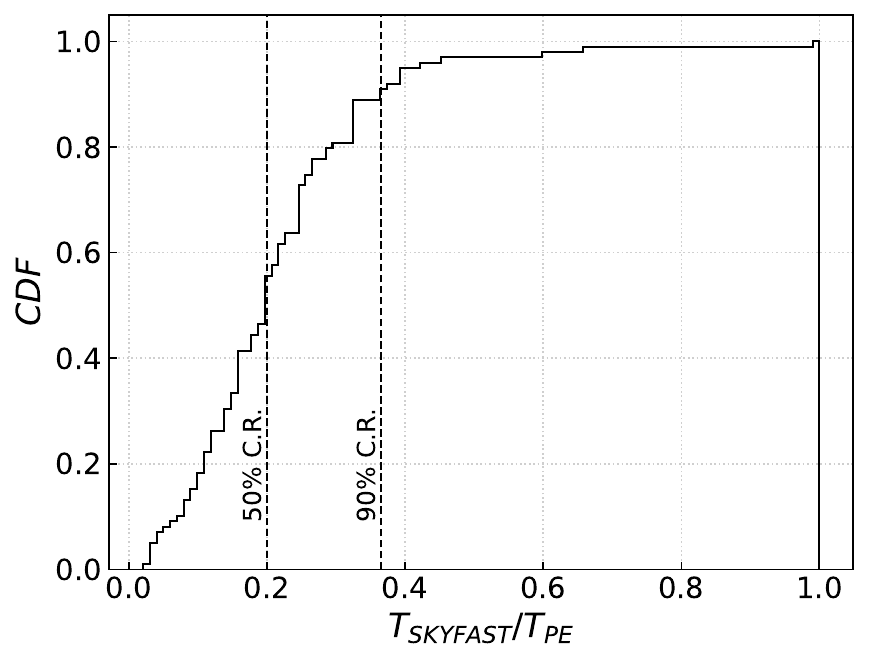}
    \caption{Cumulative distribution of the fraction of the PE runtime needed for \texttt{SKYFAST} to achieve the information entropy convergence criterion and release all its outputs: intermediate posterior reconstruction, skymap, and galaxy host ranking. The setup is the same as in Figure \ref{fig:PE/skyfast}.}
    \label{fig:time-fraction}
\end{figure}

\begin{figure}[ht]
    \centering
    \includegraphics[width = 0.4\textwidth]{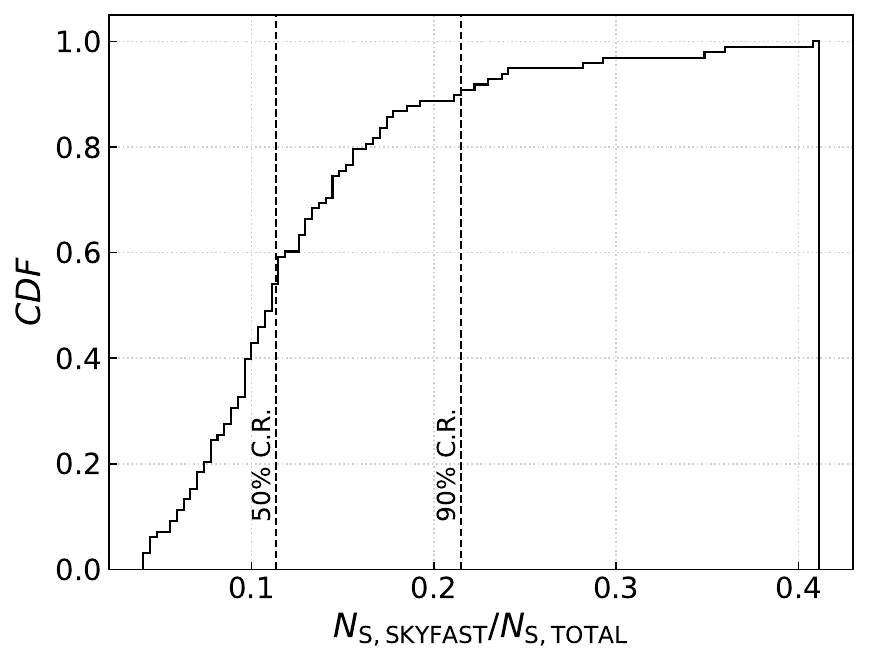}
    \caption{Cumulative distribution of the fraction of total samples needed for \texttt{SKYFAST} to achieve the information entropy convergence criterion and release the intermediate posterior reconstruction, skymap, and galaxy host ranking. The setup is the same as in Figure \ref{fig:PE/skyfast}.}
    \label{fig:sample-fraction}
\end{figure}

In Figure \ref{fig:PE/skyfast}, we compare the distribution of the total PE sampling time with the time required by \texttt{SKYFAST} to release the (intermediate) localization of the GW event. The latter is determined by the time that the PE algorithm takes to produce the necessary number of samples for \texttt{SKYFAST} to meet the convergence criterion.
Figure \ref{fig:time-fraction}, on the other hand, shows the cumulative distribution of the ratio between the \texttt{SKYFAST} and PE runtimes. 
With this specific PE pipeline and settings, \texttt{SKYFAST} completes the intermediate posterior reconstruction, skymap generation, and ranking of the most probable host galaxies for $50\%(90\%)$ of the events in less than  $\sim 20\%(36\%)$ of the total sampling time. The performance of \texttt{SKYFAST} is significantly influenced by the chosen PE pipeline and its specific configuration, as discussed further in Section \ref{sec:pipelines}.
Figure \ref{fig:sample-fraction} shows the cumulative distribution of the fraction of samples needed to achieve convergence.  \texttt{SKYFAST} requires less than $\sim 11\% (21\%)$ of the total samples generated by the PE to achieve information entropy convergence and to reconstruct the intermediate posterior for $50\% (90\%)$ of events. Contrarily to naive expectations, the distribution of the sample fraction differs from that of the total PE time fraction shown in Figure \ref{fig:time-fraction}. This difference is due to the burn-in inefficiency: during the initial burn-in period, the samplers produce correlated samples that cannot be used for analysis. 

\begin{figure}[ht]
    \centering
    \includegraphics[width = 0.4\textwidth]{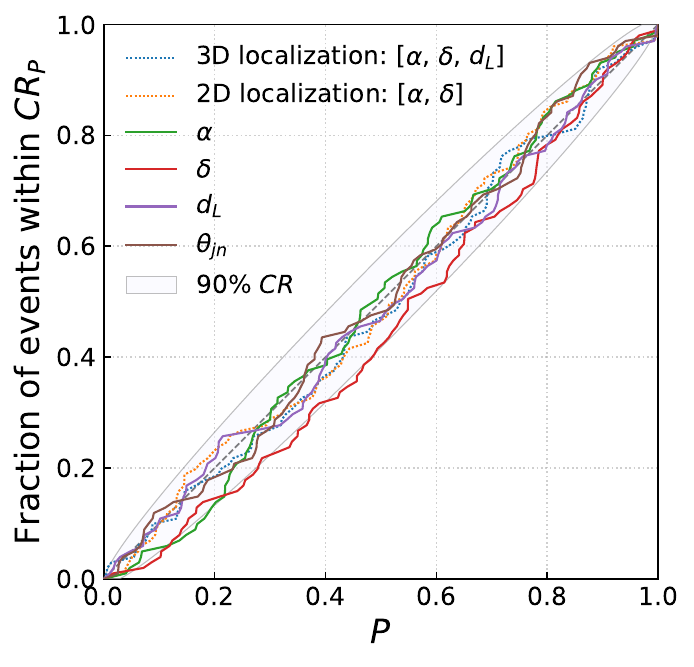}
    \caption{Probability-Probability plot (PP-plot) for the intermediate posterior distributions of $\alpha, \delta, d_L$ and $\theta_{jn}$, and for the 2D and 3D localization posterior distributions. The grey dashed diagonal line is the expected distribution, and the shaded areas 
    are the corresponding 90\% credible intervals. The setup is the same as in Figure \ref{fig:PE/skyfast}. }
    \label{fig:pp_plot_all}
\end{figure}

Lastly, we test the statistical robustness of the reconstructed intermediate posterior distribution by plotting the fraction of injected BBH events found within a credible region $\rm{CR}_P$ as a function of the encompassed probability P for the single parameters $\alpha, \, \delta, \, d_L$ and $\theta_{jn}$, and the 2D and 3D localizations. The expected distribution is indeed a diagonal line.
As shown in Figure \ref{fig:pp_plot_all}, all the probability-probability plots lie on the diagonal within the 90\% credible interval for the cumulative distribution computed from a beta distribution as in \cite{Cameron_2011}. This allows us to assess that the intermediate posteriors reconstructed with \texttt{SKYFAST} are statistically unbiased.

\subsection{Galaxy ranking}\label{sec:galaxyrank}
\begin{figure}[t!]
    \centering
    \includegraphics[width = 0.4\textwidth]{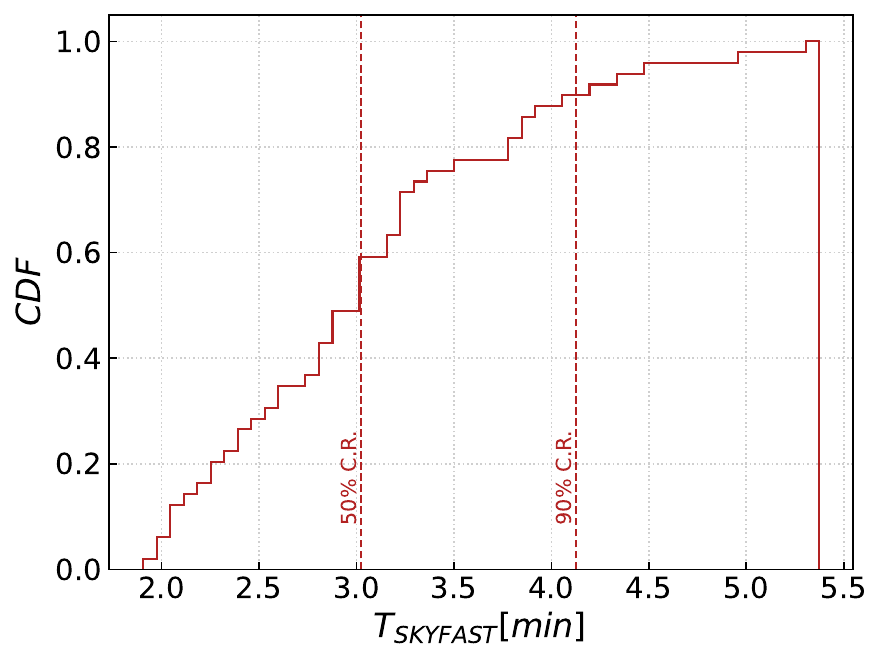}
    \caption{Cumulative distribution of time needed by \texttt{SKYFAST} to reconstruct the posterior and to produce all the relevant information.
    The times are referred to a Intel\textsuperscript{\textregistered} Xeon\textsuperscript{\textregistered} Gold 6140M processor with a 2.30GHz clock rate.
    }
    \label{fig:skyfast_time}
\end{figure}

To evaluate the performance of \texttt{SKYFAST} in ranking the most probable hosts we analyze 50 simulated BNS signals. We inject these signals into Gaussian noise using the \texttt{IMRPhenomD} waveform, following the configuration discussed in Section \ref{sec:intermediate}. The localization parameters of the BNSs are randomly extracted from the GLADE+ catalog, with a luminosity distance cut of 100 Mpc to have a reasonable number of galaxies in the localization volume. We perform the parameter estimation using \texttt{Bilby} with the \texttt{DYNESTY} sampler \cite{speagle2020dynesty}, employing the Reduced Order Quadrature (ROQ) technique \cite{morisaki:2023}. This setup closely mirrors the online parameter estimation pipeline used by the LVK Collaboration \cite{emfollow}. 
However, with this configuration, posterior samples are only available at the end of the inference due to the nested sampling scheme of \texttt{DYNESTY}. Consequently, we use \texttt{SKYFAST} directly on the final samples, thus without the intermediate posterior reconstruction. Given that the median distance of the extracted galaxies is $\sim 67$ Mpc, we obtain the following results:
\begin{itemize}
    \item the median number of galaxies in the 90\% credible volume is 30;
    \item on average, the true host galaxy is ranked as 5\textsuperscript{th}.
\end{itemize}
The distribution of the time needed for \texttt{SKYFAST} to obtain the ranked host galaxy list from the posterior samples is shown in Figure \ref{fig:skyfast_time}. Most of the events require a runtime of less than three minutes. 
The current online PE pipeline takes around 10 minutes, on average, to complete the inference for BNS events \cite{morisaki:2023}. Our results suggest that by adding a few minutes, it is possible to obtain a list of the most probable host galaxies along with an estimate of the inclination angle conditioned to the position of each galaxy in the list\footnote{All the findings presented in Section \ref{sec:validation} are obtained running \texttt{SKYFAST} on Intel\textsuperscript{\textregistered} Xeon\textsuperscript{\textregistered} Gold 6140M processors with a 2.30GHz clock rate. The runtime can be further reduced by using more powerful processors. For instance, the average time needed to run \texttt{SKYFAST} on a GW event with a few thousand samples on an Apple\textsuperscript{\textregistered} M3 Pro chip is about one minute.}.

\section{Discussion}\label{sec:discussion}

In the following, we discuss the relevance of the various applications of \texttt{SKYFAST}. In section \ref{sec:pipelines}, we comment on how the performance of \texttt{SKYFAST} depends not only on the PE pipeline from which it takes the samples but also on the specific setup within the same pipeline. In section \ref{sec:theta_em_searches}, we discuss the contribution that \texttt{SKYFAST} can make in optimizing the EM follow-up searches, with particular emphasis on the importance of the inclination angle information released with the galaxy ranking.

\subsection{Combination with different PE pipelines}\label{sec:pipelines}

The performance of \texttt{SKYFAST} depends on the parameters estimation pipeline from which it takes posterior samples. 
It should be kept in mind that the time gain distribution shown in Figure \ref{fig:time-fraction} is sensitive not only to the algorithm used for the inference, but also to the specific PE setup. For example, running the same pipeline with more (less) independent samplers has the effect of decreasing (increasing) the time gain, because of the burn-in inefficiency. 

Releasing as soon as possible an accurate sky localization along with a list of potential host galaxies is of great relevance for GW events expected to have an EM counterpart, such as binary systems containing at least one neutron star.
Currently, the online PE in the LVK Collaboration uses the ROQ technique and the \texttt{DYNESTY} sampler. With this setup and using the \texttt{IMRPhenomD} waveform, the PE wall time for BNS events is around ten minutes. As mentioned previously, accessing samples during the run is not possible with a nested sampler like \texttt{DYNESTY}. However, using \texttt{SKYFAST} in combination with an MCMC algorithm such as \texttt{Bilby-MCMC} to generate an intermediate skymap would require more time than the average wall time of the current online PE pipeline. This is due to \texttt{Bilby-MCMC} being less optimized compared to \texttt{DYNESTY}, mainly since distance marginalization cannot be used for our specific purposes.
Nonetheless, \texttt{SKYFAST} can still come into play immediately after the online PE process ends. By adding just a few minutes to the PE runtime, it becomes possible to generate an analytical reconstruction of the posterior using \texttt{SKYFAST} with all the samples produced by the PE. This posterior can then be used to generate a list of the most probable galaxy hosts within the GW event localization region, along with the associated estimate of $\theta_{jn}$, as discussed in Section \ref{sec: ex_summary}.

The full potential of \texttt{SKYFAST} can be realized when used in combination with Hamiltonian Monte Carlo (HMC) methods \cite{neal2012mcmc}, which sample the posterior distribution more efficiently than traditional MCMC or Nested Sampling methods. However, a significant drawback of HMC algorithms is their requirement to compute the gradient of the likelihood, and consequently the gradient of the waveform. This task is computationally expensive and could potentially make the entire pipeline inefficient.
Given the ongoing efforts in developing differentiable waveforms \cite{Margossian_2019,Edwards:2023sak, Schmidt:2020yuu}, we expect HMC to become the standard PE strategy in the future. Combining \texttt{SKYFAST} with HMC would further reduce the time required to obtain accurate localization and galaxy host information.

\subsection{Galaxy ranking and inclination angle}\label{sec:theta_em_searches}

As discussed in the previous section, one potential output of \texttt{SKYFAST} is the generation of a ranked list of galaxies within a given credible volume, typically 90\%, based on their probability of being the host. 
If the pipeline does not provide access to the samples during the run, \texttt{SKYFAST} can be run right after the inference is complete, generating the ranked list within minutes.

\begin{figure}[t!]
    \centering
    \includegraphics[width = 0.49\textwidth]{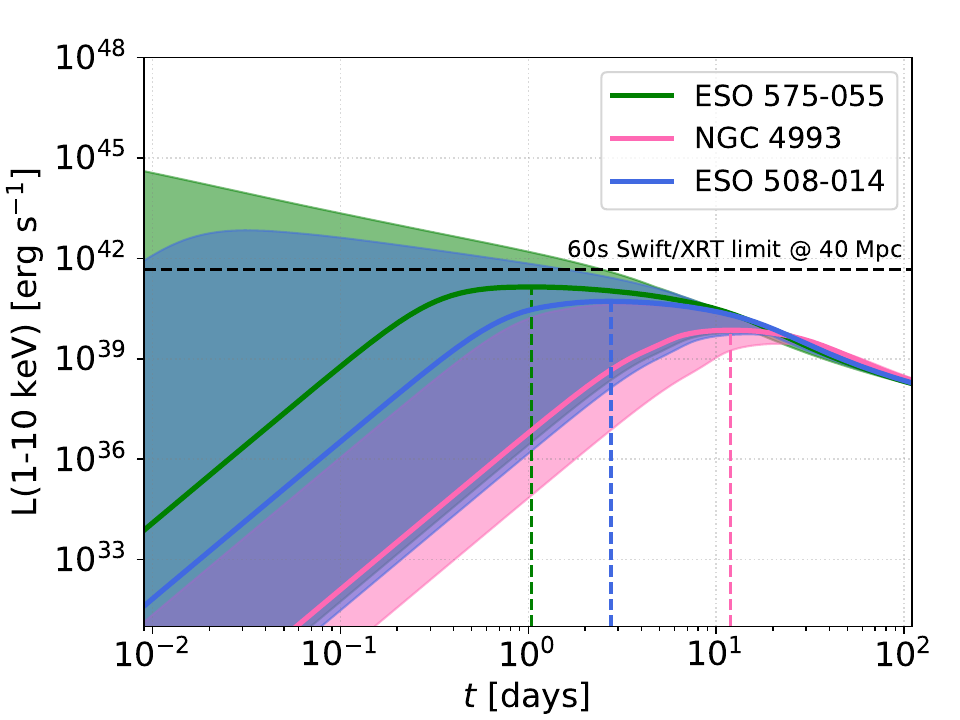}
    \caption{Impact of different conditioned inclination angle posteriors on the expected GRB afterglow light curve in the X-ray band. The solid lines show the expected luminosity curve for the median inclination angle of each galaxy, as reported in Table \ref{tab:gw170817_list}. The shaded areas represent the uncertainty in the luminosity curve, reflecting the 90\% credible intervals of the inclination angle. The colored dashed lines indicate the times of the luminosity peaks. We report the 60-second sensitivity limit of \textit{Swift}/XRT \cite{SwiftScience:2004ykd,Evans:2016dgd} at a distance of 40 Mpc, which is comparable to the distances of all potential hosts of GW170817.
    }
    \label{fig:lightcurves}
\end{figure}

A feature that differentiates \texttt{SKYFAST} from other galaxy ranking tools \cite{Salmon:2019bqy, ned_gwf,Arcavi:2017vbi,Ducoin:2019rdv} is the possibility to compute the inclination angle posterior conditioned to the position of each galaxy in the list. Currently, LVK GCN notices and circulars do not include inclination angle information. However, we do integrate this parameter into \texttt{SKYFAST} posterior reconstruction because it plays a crucial role in optimizing the EM follow-up campaign. Specifically, the inclination angle of the binary system relative to our line of sight affects the multi-wavelength light curve of possible EM counterparts such as, e.g., GRBs.
To further illustrate this point, we return to our working example, GW170817/GRB 170817A. In Figure \ref{fig:lightcurves}, we plot the expected GRB afterglow light curves in the X-ray band, assuming the event is located in each of the first three galaxies listed in Table \ref{tab:gw170817_list}. We compute the light curves using \texttt{afterglowpy} \cite{Ryan:2019fhz}, and assuming a Gaussian profile for the structured GRB jet. We adopted the best-fit GRB parameters obtained from \cite{Ryan:2019fhz} for GW170817 (see their Table 3). However, we use the luminosity distance and inclination angle values reported for each galaxy in Table \ref{tab:gw170817_list}. The shaded areas represent the uncertainties in the luminosity curves associated with the 90\% credible intervals of the inclination angle estimate. 
The different inclination angle posteriors obtained by conditioning on the position of each galaxy have a significant impact on the afterglow peak time and peak luminosity. Vertical lines indicates the expected times of the luminosity peak for each curve, revealing an order of magnitude discrepancy mainly driven by variations in the inclination angle posteriors among galaxies.
For certain configurations (such as ESO 575-055 and partially for ESO 508-014), the smallest inclination angles within the 90\% credible intervals are comparable to the half-width of the jet core. This results in a time-decreasing light curve typically observed for GRBs viewed on-axis. 
The shape of the light curve affects the detectability of the EM emission. As an example, we compare the simulated X-ray light curves with the limiting luminosity that can be reached by \textit{Swift}/XRT, considering an exposure time of 60 s and a distance of 40 Mpc. It can be seen that the afterglow could be detected until approximately 1.5 (2.5) days after the merger if the host galaxy is ESO 575-055 or ESO 508-014, respectively. To potentially detect the afterglow emission if the host galaxy is NGC 4993, a much longer exposure time is needed. 

Therefore, the conditioned inclination angle posteriors, along with a GRB jet model and the sensitivity limits of various instruments at the distance of the potential host galaxy, play a crucial role in determining whether and when the afterglow emission can be observed with a specific EM facility, depending on the host candidate this is also key to optimize the observational strategy in terms of exposure time of the various tilings, to maximize the chance of detection of a possible EM counterpart. 
As a consequence, \texttt{SKYFAST} can significantly contribute to optimizing the EM follow-up campaign by providing essential information to design a tiling strategy tailored to each potential host galaxy in the list. 
Finally, the information on the inclination angle could be used to estimate the likelihood that a transient source detected during the EM follow-up campaign is the counterpart to (or unrelated to) a GW event.

\section{Conclusions}\label{sec:conclusion}

We presented \texttt{SKYFAST}, a new tool for the rapid localization of GW events and the ranking of the most probable galaxy hosts. \texttt{SKYFAST} uses Bayesian non-parametrics to reconstruct an analytical posterior distribution of sky position, luminosity distance and inclination angle from the posterior samples produced by a PE pipeline. We developed \texttt{SKYFAST} to run in parallel with a PE pipeline and update the reconstructed posterior as new samples are produced, until the information entropy reaches a plateau. In general, \texttt{SKYFAST} needs less samples than a full PE to reconstruct an accurate posterior for the parameters of interest. This feature results in relevant time gains that can be crucial for the prompt identification of the EM counterpart. 

Together with the reconstructed posterior, \texttt{SKYFAST} releases a list of all the galaxies from a catalog of choice that are contained within the 90\% credible volume of the GW event, ranked by their probability of being the host. In this work, we used the GLADE+ catalog as an example, but the algorithm has the great advantage of being easily interfaced with any galaxy catalog. We tested \texttt{SKYFAST} on the posterior samples of GW170817. We found that the reconstructed posterior is in agreement with the final PE results and the sky localization area is also compatible with the one reported in the literature. We ranked all the galaxies in the 90\% credible volume, with the true host, NGC 4993, being the third most probable one.
Thanks to the analytical form of the reconstructed posterior, \texttt{SKYFAST} can easily infer the inclination angle posterior conditioned on the position of each galaxy in the list. This not only breaks the degeneracy between $d_L$ and $\theta_{jn}$, resulting in a more stringent inclination angle constraint, but it also allows for more precise predictions of the GRB afterglow observed light curves expected if the GW event were located in each potential galaxy host. Interestingly, we found that the light curves can vary significantly among galaxies, mainly due to variations in the conditioned inclination angle posterior.
The conditioned inclination angle posterior, combined with the sensitivity of different instruments at the distance of potential hosts, is crucial for predicting whether and when each instrument can detect GRB afterglow emission. This depends on the galaxy it is pointing to and helps choose an optimal exposure time to maximize the chance of detection.

We tested the statistical robustness of \texttt{SKYFAST} by running it on a population of 100 simulated BBH events. We built PP-plots for the posteriors of all the reconstructed parameters, as well as for the 2D and 3D localizations. We found that the reconstructed posterior is unbiased and accurately describes the distribution from which the input samples are taken. Additionally, we found that \texttt{SKYFAST} is able to release the reconstructed posterior within one-fifth (one third) of the total PE runtime for 50\% (90\%) of the GW events, using less than one-tenth (one fifth) of the samples. However, we note that the exact values of the time and sample fractions depend on the PE pipeline used and its specific setup (e.g., the number of independent samplers and parallelization scheme).
To test the performance of \texttt{SKYFAST} in ranking the most probable galaxy hosts, we ran it on 50 simulated BNS events. Specifically, we produced the input samples with \texttt{Bilby}, using a nested sampling scheme accelerated with the ROQ technique. With this setup, the PE runtime is 8 minutes, while the \texttt{SKYFAST} runtime is 1-2 minutes on average. These times depend on the computational power of the machines used and can be further reduced with next-generation processors. Regarding galaxy ranking, given that the median luminosity distance of our BNSs was around 70 Mpc, we found an average of 30 galaxies in the 90\% credible volume, with the true host in the 5th position.

In conclusion, \texttt{SKYFAST} is a lightweight and user-friendly tool designed to reconstruct analytical posteriors for sky localization, luminosity distance, and inclination angle from posterior samples generated by a PE pipeline, achieving these tasks in a fraction of the PE runtime. Even if real-time sample access is unavailable during PE, \texttt{SKYFAST} can be launched immediately after PE completion to provide a ranked list of potential galaxy hosts within just one or two minutes. A novel feature of \texttt{SKYFAST} compared to other tools is the inclusion of inclination angle information, crucial for astronomers to plan optimized EM follow-up strategies. Lastly, \texttt{SKYFAST} can be used offline for various purposes, such as cross-correlating GW events with different galaxy catalogs to infer values of $H_{0}$ and other cosmological parameters.

\section*{Acknowledgement}

We thank Stefano Rinaldi for providing the scratch from which \texttt{SKYFAST} was developed and for his assistance with \texttt{FIGARO}. We thank Gianluca Maria Guidi and Francesco Pannarale for useful discussions and for carefully reading the manuscript. This material is based upon work supported by NSF's LIGO Laboratory which is a major facility fully funded by the National Science Foundation.
This work has been supported by the project BIGA - ``Boosting Inference for Gravitational-wave Astrophysics" funded by the MUR Progetti di Ricerca di Rilevante Interesse Nazionale (PRIN) Bando 2022 - grant 20228TLHPE - CUP I53D23000630006. GD acknowledges financial support from the National Recovery and Resilience Plan (PNRR), Mission 4 Component 2 Investment 1.4 - National Center for HPC, Big Data and Quantum Computing - funded by the European Union - NextGenerationEU - CUP B83C22002830001. AR acknowledges financial support from the Supporting TAlent in ReSearch@University of Padova (STARS@UNIPD) for the project ``Constraining Cosmology and Astrophysics with Gravitational Waves, Cosmic Microwave Background and Large-Scale Structure cross-correlations''

\bibliography{bibliography}
\bibliographystyle{apsrev4-1}

\end{document}